\documentclass{article}
\usepackage{graphicx} 

\title{An Annotated Glossary for Data Commons, Data Meshes, 
and Other Data Platforms}

\author{Robert L. Grossman\protect\\
University of Chicago \protect\\
and Open Commons Consortium
}
\date{February 26, 2021 \protect\\
Revised 2021--2024 \protect\\
Version 1.07 (April 22, 2024)
}

\usepackage[utf8]{inputenc}

\usepackage{graphicx}
\usepackage[margin=1.25in]{geometry}
\usepackage{hyperref}
\usepackage{enumitem}

\newcommand{\ph}[1]{\medbreak \noindent {\bf #1}}

\begin{document}

\maketitle

\section*{Abstract}
Cloud-based data commons, data meshes, data hubs, and other data platforms are important ways to manage, analyze and share data to accelerate research and to support reproducible research.  This is an annotated glossary of some of the more common terms used in articles and discussions about these platforms.  

\section{Introduction}

Cloud-based data commons, data meshes, data hubs, and other data platforms are important ways to manage, analyze and share data to accelerate research and to support reproducible research.  This is an annotated glossary of some of the more common terms used in articles and discussions about these platforms.  Definitions are from \cite{grossman_progress_2018}, \cite{grossman_data_2019}, or \cite{grossman2024safe}, or as otherwise indicated.

Cloud-based data commons have been around for over a decade \cite{grossman2023ten}, and this glossary reflects a broad community consensus about the terms associated with data commons.  On the other hand, data meshes and data fabric are emerging architectures (at least when this glossary was last updated in 2024) for managing data within an organization, and there is not yet a consensus on their definitions.  The definitions here around data meshes, data fabrics, and associated concepts should be thought of as working definitions that are likely to change in the future.

\section{Glossary}

\ph{Application Programmer Interfaces (API).} An An application programming interface (API) is a type of software interface and provides a way for two or more computer programs or software components to communicate with each other. Said another way, APIs are software-based intermediaries that provide a way for two software components to exchange data.

\ph{Authorized environment (aka authorized analysis environment).}  An authorized environment is a cloud platform workspace or computing infrastructure that is approved for the use or analysis of controlled access data \cite{grossman2024safe}.

\ph{Authorized platform identifier (APID).}  The SAFE Framework \cite{grossman2024safe} assumes that cloud platforms have a globally unique identifier (GUID) identifying them called the authorized platform identifier (APID).

\ph{Authorized platform network (APN) and authorized platform network identifier (APNI).}  The SAFE Framework \cite{grossman2024safe} assumes that cloud platforms form networks consisting of two or more cloud platforms called authorized platform networks (APN) that are identified by a globally unique identifier called an authorized platform network identifier (APNI).

\ph{Authorized user.}  An authorized user is someone who has applied and been approved for access to controlled-access data.

\ph{Authorized region ID.}  The SAFE Framework \cite{grossman2024safe} assumes that geographic regions are identified by a globally unique identifier called an Authorized Region ID (ARID).

\ph{Cloud computing.} Cloud computing is a model for enabling ubiquitous, convenient, on-demand network access to a shared pool of configurable computing resources (e.g., networks, servers, storage, applications, and services) that can be rapidly provisioned and released with minimal management effort or service provider interaction \cite{mell_nist_2011}.

Public cloud service providers provide on-demand delivery of IT resources through the Internet or private networks with a pay-as-you-go pricing on an as-needed basis. Private cloud service providers set up and operate their own cloud computing infrastructure called private clouds for their own organizations.

\ph{Cloud service provider (CSP).} See cloud computing. 

\ph{Commons.} A commons is a natural, cultural or digital resource accessible to all members of a community, or more broadly of a society. Examples include a pasture for animals to graze in a village, a dog park for dogs in a city neighborhood, or natural materials such as air, water for society in general. These resources are held in common, through a partnership, a not-for-profit, or other entity, but not owned privately for commercial gain \cite{ostrom_governing_1990}.

A particularly important type of commons are digital commons containing data. A specific type of digital commons is a data commons, which are used by projects and communities to create open digital resources to accelerate the rate of discovery and increase the impact and the benefits of the data they hold. More formally, data commons are software platforms that co-locate: 1) data, 2) cloud-based computing infrastructure, and 3) commonly used software applications, tools and services to create a resource for managing, analyzing, integrating and sharing data with a community.

\ph{Common Services Operations Center (CSOC).}  A Common Services Operations Center  is an operations center operated by a commons services provider for setting up, configuring, operating, and monitoring data commons, data meshes, data hubs, and other data platforms for managing, analyzing, and sharing data.

\ph{Commons Services Provider.}  A Common Services Provider is an organization that provides services for setting up, configuring, operating, and monitoring data commons, data meshes, data hubs, and other data platforms for managing, analyzing, and sharing data.

\ph{Container.}  A standard unit of software that packages up code and all its dependencies so the application runs quickly and reliably from one computing environment to another (for example, Docker).

\ph{Data cleaning.} By data cleaning, we mean identifying and correcting problems with data formats, allowed data values, required documentation, etc. of submitted data.  

\ph{Data clouds.} A data cloud is a cloud computing platform for managing, analyzing and sharing datasets.

\ph{Data commons.}  A data commons co-locates data with cloud computing infrastructure and commonly used software services, tools, and applications for managing, integrating, analyzing and sharing data that are exposed through web portals and APIs to create an interoperable resource for a research community.  A data commons provides services so that the data is findable, accessible, interoperable and reusable (FAIR) \cite{wilkinson_fair_2016}.

\ph{Data commons operating models.}   A standard operating model for a platform is to separate the platform owner, the platform developer and the platform operator \cite{van_alstyne_pipelines_2016}.  This model can be adapted for data commons and related data platforms by introducing some additional roles, such as: data platform sponsor / owner; cloud service provider; commons services provider; data platform operator; and, data platform developer.  In this way, the platform operator role is split between the operator of the cloud services (such as AWS, GCP or Azure); the operator of the commons services; and, the operator of the data platform itself, which relies on the cloud services provider and commons services provider.

\ph{Data curation.}  We define data curation to be the process of identifying a data model for a study and aligning submitted data to the agreed upon data model.

\ph{Data ecosystems aka data meshes.}  See data meshes.  Data meshes in the context of this article contain multiple data commons, data repositories, cloud computing infrastructure, cloud based resources, including databases and  knowledgebases, and other applications that can interoperate using a common set of services (sometimes called mesh services or framework services) for authentication, authorization, accessing data,  analyzing data and related functions.  

\ph{Data fabric.}  Data meshes and data fabrics are emerging architectures for managing data within an organization.  The goal of both architectures  is to improve data management, but they do so through different approaches and with distinct focuses. A data mesh emphasizes decentralization and domain-specific ownership of data by subject matter experts. A data fabric focuses on providing a unified, consistent and integrated approach to data throughout an organization.  A data mesh is a bottom up approach and a data fabric is a top down approach to data management.

\ph{Data hub.}  A data hub is a data platform in a data mesh that supports search and discovery of data in 1, 2 or more data platforms in a data mesh.  Typically,  the data itself remains in the data repository, data commons, or other data platform in the mesh, but the metadata for the data is typically replicated in the data hub.  Data hubs often are connected to workspaces or analysis environments operated by the data hub so the data can be analyzed.

\ph{Data lake.} A data lake is a system for storing data as objects, where the objects have an associated GUID and (object) metadata, but there is no data model for interpreting the data within the object.

\ph{Data governance.}  Data governance is the set of roles, processes, policies, standards and tools that promote the availability, quality, security and privacy of an organization's data.  In the context of data platforms for sharing research data, data governance includes the agreements and processes for contributing data to a data platform, the agreements and processes for approving users to access and analyze data on the data platform, the agreements and processes for redistributing data to other platforms for analysis, and related activities.    

\ph{Data harmonization.}  1)  Data harmonization as the process that brings together data from multiples sources and applies uniform and consistent processes, such as uniform quality control metrics to the accepted data; mapping the data to a common data model; processing the data with common bioinformatics pipelines; and post-posting the data using common quality control metrics.

2) More precisely, you can separate data harmonization into two phases.   Data harmonization of submitted data is the process of processing data from different sources and studies and curating and aligning the data to a common data model. Data harmonization of derived data is the result of processing harmonized data using a common, uniform set of data processing pipelines versus each site separately processed data with their own set of pipelines that may have minor or major differences.

\ph{Data meshes (aka data ecosystems).} In the context of this article, data meshes contain multiple data commons, data repositories, cloud computing infrastructure, cloud based resources, including databases and  knowledgebases, and other applications that can interoperate using a common set of services (sometimes called mesh services or framework services) for authentication, authorization, accessing data,  analyzing data and related functions.  

More generaly, both data meshes and data fabrics are emerging architectures for managing data within an organization. Quoting from the glossary entry for data fabrics: The goal of both architectures is to improve data management, but they do so through different approaches and with distinct focuses. A data mesh emphasizes decentralization and domain-specific ownership of data by subject matter experts. A data fabric focuses on providing a unified, consistent and integrated approach to data throughout an organization.  A data mesh is a bottom up approach and a data fabric is a top down approach to data management.

\ph{Data mesh services (aka data framework services).}   A core set of software services (an example of microservices) or authentication, authorization, data access, metadata access and other services necessary to support and interoperate data commons, data meshes, data hubs, and workspaces. In an influential 1984 paper called ``End-to-end arguments in systems design \cite{saltzer1984end},'' Jerome Saltzer, David Reed and David Clark argued for using as few software services as possible when building large complex distributed systems.  In the 2018 publication \cite{grossman_proposed_2018}, Grossman proposed a similar approach called the {\em narrow middle architecture} for data platforms supporting scientific and biomedical data.  Data mesh services adhere to this principle.

\ph{Data object.} In cloud computing, a data object consists of data, a key, and associated metadata. The data can be retrieved using the key and the metadata associated with a specific data object can be retrieved, but more general queries are not supported.  Amazon’s S3 storage system is a widely used storage system for data objects. Compare to structured data.

\ph{Data platform.} A data platform is a type of software platform [4] that brings together: i) data contributors that provide data to the platform; ii) data consumers that access and use data from the platform; iii) a platform owner that is responsible for funding the platform and the data and platform governance;  iv) a platform developer that develops the platform; and a v) platform operator that operates the platform.  

\ph{Digital Object Identifier (DOI).} A Digital Object Identifier  is an identifier used to permanently and stably identify (usually digital) objects. DOIs provide a standard mechanism for retrieval of metadata about the object, and generally a means to access the data object itself.  A DOI is a specific ISO standard for a particular class of persistent identifiers and is supported by the DOI Foundation. 

\ph{Data portal.} A data portal is a website that provides interactive access to data in an underlying data management systems, such as a database.  Data commons, data lakes can also have data portals. 

\ph{FAIR data.} FAIR data are data which meet the principles of findability, accessibility, interoperability, and reusability \cite{wilkinson_fair_2016}.  There is now an extensive literature on FAIR data .

\ph{Framework services.} Framework Services or Data Commons Framework (DCF) Services is the term used by Gen3 \cite{gen3_welcome_2024} to refer to data mesh services in the narrow middle architecture, for data meshes, such as the NCI Cancer Research Data Commons \cite{wang2024nci}.  These are set of standards-based services with open APIs for authentication, authorization, creating and accessing FAIR data objects, and for working with bulk structured data in machine readable, self-contained format.  

\ph{Globally Unique Identifier (GUID).} A GUID is an essentially unique identifier that is generated by an algorithm so that no central authority is needed, but rather different programs running in different locations can generate GUID with a low probability that they will collide.  A common format for a GUID is the hexadecimal representation of a 128 bit binary number.  

\ph{Microservices.}  Microservices are a software architecture that organizes software into small, independent services that communicate over well defined APIs.  These services can be developed, set up, and scaled independently.  A more traditional architecture is to put all the  APIs and other required functionality into a single application. This is sometimes called a monolithic architecture.  Microservices provide important advantages for large-scale systems that require scalability and must continue to evolve even as their code base grows very large, but increases the complexity of operating small-scale systems.

\ph{Narrow Middle Architecture.}   The narrow architecture is an architecture that applies the end-to-end argument in systems design  principle (see the glossary entry) to  data commons and data meshes \cite{grossman_proposed_2018}.   With this approach, the fewest possible services are used for the core of the system ({\em the narrow middle}) and only these services are standardized. With this architecture, the importing, curation and integration services for getting data into the commons (one ``end'') and the data exploration, analysis and collaboration services for getting knowledge out of the commons (the other ``end'') are not standardized, but instead are the focus of different competing and innovative efforts until the community begins to recognize and adopt approaches that seem to be most effective. 

\ph{Persistent Identifier (PID).}  A persistent identifier (PI or PID) is a long-lasting reference to a document, file, web page, or other digital object. Most PIDs have a unique identifier which is linked to the current address of the digital object or metadata associated with the digital..

A Digital Object Identifier (DOI) is a type of a persistent identifier that is an implementation of the handle architecture and is standardized by the International Organization for Standardization (ISO).   You can turn any DOI starting with 10 into a URL by adding https://doi.org/ before the DOI as in  https://doi.org/10.1038/s41597-023-02029-x

A Data GUID is a type of persistent identifier designed to support data objects and standardized by the GA4GH and the Open Commons Consortium.

\ph{Platform (aka software platform).}  Also, see data platform.  A platform is an ecosystem that has four main types of participants \cite{van_alstyne_pipelines_2016}. Consumers that use the product or service provided by the platform.  Producers that provide the product or service.  For example, with a data platform (see above), data contributors are producers and data users (i.e. researchers) are the consumers.  The platform operator operates the platform. The platform sponsor (aka platform owner) that is responsible for funding and governing the platform (both data and platform governance), and managing the associated IP.  

\ph{Platform Governance.} Platform governance includes approving cloud platforms as having the right to distribute datasets to other platforms and to authorized users on the platforms and approving cloud platforms as authorized environments so that the cloud platforms can be used by authorized users to access, analyze, and explore datasets

\ph{Platform operator.}  A platform operator is responsible for operating cloud platforms that provide access to datasets and to analysis tools and implementing controls that balance the availability of data with the obligation to protect the confidentiality and security of the data.

(1) There is an agreement between the Project Sponsor and a Platform Operator that sets the terms and conditions that enable the cloud platform to distribute data to authorized users and to transfer the data to authorized platforms (Right to Distribute Data Agreement). (2) There is an agreement between the Project Sponsor and a Platform Operator that sets the terms and conditions that enable the cloud platform to provide an environment to users so they can access and analyze data (Authorized Environment Agreement).

\ph{Structured data.} Data is structured if it is organized into records and fields, with each record consisting of one or more data elements (data fields).   In biomedical data, data fields are often restricted to controlled vocabularies to make querying them easier. 

\ph{Workspace.}  A workspace is another name for a cloud-based environment for analyzing data (aka an authorized analysis environment).


\begin{thebibliography}{10}

\bibitem{gen3_welcome_2024}
{Center for Translational Data Science, University of Chicago}.
\newblock Welcome to {Gen3}.
\newblock https://gen3.org.

\bibitem{grossman_proposed_2018}
Robert Grossman.
\newblock A {Proposed} {End}-{To}-{End} {Principle} for {Data} {Commons}. 
\newblock Medium, July 6, 2018.
\newblock Retrieved from \url{https://medium.com/@rgrossman1/a-proposed-end-to-end-principle-for-data-commons-5872f2fa8a47}

\bibitem{grossman_progress_2018}
Robert~L. Grossman.
\newblock Progress {Toward} {Cancer} {Data} {Ecosystems}.
\newblock {\em The Cancer Journal}, 24(3):126--130, May 2018.

\bibitem{grossman_data_2019}
Robert~L. Grossman.
\newblock Data {Lakes}, {Clouds}, and {Commons}: {A} {Review} of {Platforms} for {Analyzing} and {Sharing} {Genomic} {Data}.
\newblock {\em Trends in Genetics}, 35(3):223--234, March 2019.

\bibitem{grossman2023ten}
Robert~L Grossman.
\newblock Ten lessons for data sharing with a data commons.
\newblock {\em Scientific Data}, 10(1):120, 2023.

\bibitem{grossman2024safe}
Robert~L. Grossman, Rebecca~R. Boyles, Brandi~N. Davis-Dusenbery, Amanda Haddock, Allison~P. Heath, Brian~D. O’Connor, Adam~C. Resnick, Deanne~M. Taylor, and Stan Ahalt.
\newblock A {Framework} for the {Interoperability} of {Cloud} {Platforms}: {Towards} {FAIR} {Data} in {SAFE} {Environments}.
\newblock {\em Scientific Data}, 11(1):241, February 2024.

\bibitem{mell_nist_2011}
P~M Mell and T~Grance.
\newblock The {NIST} definition of cloud computing.
\newblock Technical Report NIST SP 800-145, National Institute of Standards and Technology, Gaithersburg, MD, 2011.
\newblock Edition: 0.

\bibitem{ostrom_governing_1990}
Elinor Ostrom.
\newblock {\em Governing the commons: {The} evolution of institutions for collective action}.
\newblock Cambridge university press, 1990.

\bibitem{saltzer1984end}
Jerome~H Saltzer, David~P Reed, and David~D Clark.
\newblock End-to-end arguments in system design.
\newblock {\em ACM Transactions on Computer Systems (TOCS)}, 2(4):277--288, 1984.

\bibitem{van_alstyne_pipelines_2016}
Marshall~W. Van~Alstyne, Geoffrey~G. Parker, and Sangeet~Paul Choudary.
\newblock Pipelines, platforms, and the new rules of strategy.
\newblock {\em Harvard business review}, 94(4):54--62, 2016.

\bibitem{wang2024nci}
Zhining Wang, Tanja~M Davidsen, Gina~R Kuffel, KanakaDurga Addepalli, Amanda Bell, Esmeralda Casas-Silva, Hayley Dingerdissen, Keyvan Farahani, Andrey Fedorov, Sharon Gaheen, et~al.
\newblock Nci cancer research data commons: Resources to share key cancer data.
\newblock {\em Cancer Research}, 2024.

\bibitem{wilkinson_fair_2016}
Mark~D. Wilkinson, Michel Dumontier, IJsbrand~Jan Aalbersberg, Gabrielle Appleton, Myles Axton, Arie Baak, Niklas Blomberg, Jan-Willem Boiten, Luiz~Bonino da~Silva~Santos, Philip~E. Bourne, Jildau Bouwman, Anthony~J. Brookes, Tim Clark, Mercè Crosas, Ingrid Dillo, Olivier Dumon, Scott Edmunds, Chris~T. Evelo, Richard Finkers, Alejandra Gonzalez-Beltran, Alasdair~J.G. Gray, Paul Groth, Carole Goble, Jeffrey~S. Grethe, Jaap Heringa, Peter~A.C ’t Hoen, Rob Hooft, Tobias Kuhn, Ruben Kok, Joost Kok, Scott~J. Lusher, Maryann~E. Martone, Albert Mons, Abel~L. Packer, Bengt Persson, Philippe Rocca-Serra, Marco Roos, Rene van Schaik, Susanna-Assunta Sansone, Erik Schultes, Thierry Sengstag, Ted Slater, George Strawn, Morris~A. Swertz, Mark Thompson, Johan van~der Lei, Erik van Mulligen, Jan Velterop, Andra Waagmeester, Peter Wittenburg, Katherine Wolstencroft, Jun Zhao, and Barend Mons.
\newblock The {FAIR} {Guiding} {Principles} for scientific data management and stewardship.
\newblock {\em Scientific Data}, 3(1):160018, December 2016.

\end{thebibliography}

\end{document}